\documentclass[reprint,secnumarabic,amssymb,nobibnotes,aps,pra,superscriptaddress]{revtex4-2}

\newcommand{\tsp}[1]{\textsuperscript{#1}}
\usepackage{diagbox}
\usepackage[none]{hyphenat}
\usepackage{graphicx}
\usepackage{xcolor}
\usepackage{scalerel}
\usepackage{mathtools} 
\usepackage{amsmath}
\usepackage{physics}
\usepackage{booktabs}
\usepackage{multirow}
\usepackage{tablefootnote}
\definecolor{bluecitation}{RGB}{45, 48, 146}
\usepackage[colorlinks=true,allcolors=bluecitation]{hyperref}%

\newcommand\wir{$\omega_\mathrm{IR}$}

\newcommand\sesp{$4s^{-1}7p$} 
\newcommand\fsfp{$4s^{2}4p^{4}5s5p$} 

\newcommand\sess{$4s^{-1}6s$} 

\newcommand\sestates{$4s^{-1}np$} 
\newcommand\destates{$4p^{4}n{\ell}n^{\prime}\ell^{\prime}$} 

\begin{document}


\title{Probing coherent electronic superpositions of singly- and doubly-excited states of krypton with attosecond four-wave mixing spectroscopy}


\author{S. Yanez-Pagans}%
\affiliation{Department of Physics, University of Arizona, Tucson, Arizona 85721, USA}

\author{M. A. Alarc\'on}%
\affiliation{Department of Physics, University of Arizona, Tucson, Arizona 85721, USA}
\affiliation{Department of Physics and Astronomy, Purdue University, West Lafayette, Indiana 47907, USA}

\author{C. H. Greene}%
\affiliation{Department of Physics and Astronomy, Purdue University, West Lafayette, Indiana 47907, USA}

\author{A. Sandhu}%
\email{asandhu6@asu.edu}
\affiliation{Department of Physics, University of Arizona, Tucson, Arizona 85721, USA}
\affiliation{College of Optical Sciences, University of Arizona, Tucson, Arizona 85721, USA}
\affiliation{Department of Physics and CXFEL Laboratory, Arizona State University, Tempe, AZ, 85287, USA}

\date{\today}%


\begin{abstract}
Radiative nonlinear four-wave mixing can monitor the evolution of electronic wave packets, providing access to lifetimes and quantifies the light induced couplings between excited states. We report the observation of quantum beats in an autoionizing electronic wavepacket in krypton, probed using this technique. Analysis of the signal reveals that these beats originate from the contribution of previously unassigned, doubly excited states interacting with singly excited ones. We introduce a minimal theoretical model, based on multichannel quantum defect theory, that quantitatively reproduces both the wavepacket dynamics and the static spectrum. This work combines a versatile, background-free experimental scheme with a tractable model, establishing a powerful approach for the metrology and control of complex, correlated electronic states.


\end{abstract}

\maketitle


\section{Introduction}\label{sec:intro}

Ultrafast intense lasers pulses are ideally suited for the investigation of nonlinear light-matter interactions~\cite{mukamel1999principles,hamm2011concepts,boyd2020nonlinear}. In particular, third-order nonlinearities give rise to four-wave mixing (FWM) emissions, which are the lowest-order wave-mixing processes in centrosymmetric media~\cite{boyd2020nonlinear}. Due to their ability to capture information about the light-induced couplings between different electronic states and their relatively large cross-sections in comparison to other higher-order processes~\cite{zhang2009MWM}, FWM has been explored extensively since the advent of lasers. 

In recent years, FWM spectroscopy has been extended to the extreme-ultraviolet (XUV) energy regime and attosecond temporal resolutions. Typically, XUV photons are obtained through high-harmonic generation (HHG) process ~\mbox{\cite{corkum1993plasma,HHG1996,HHG2001}} and then mixed with the infrared (IR) laser photons to produce new emissions that probe the excited states of atomic and molecular systems. In terms of technique, FWM is a simple extension of attosecond transient absorption spectroscopy~\cite{Pollard1992ATAS,Goulielmakis2010ATAS,Wu2016TheoryATAS,liao2015,liao2016,liao2017,liao2017_2,Harkema:18,Harkema2021,Yanez-Pagans2022}. The XUV-IR FWM spectroscopy allows researchers to study coherent electron dynamics in time and energy resolved fashion, and is therefore a powerful tool to study the evolution of complex polyelectronic systems.

Many-electron systems exhibit autoionizing states (AISs)---above-threshold resonances with asymmetric cross-section profiles \cite{Fano1961EffectsShifts,Fano1965LineGases}--- which can act as intermediates in the implementation of optical control in the continuum. For example, a resonant dressing laser field~\cite{Yanez-Pagans2022} coupling two autoionizing states can give rise to entangled light-matter states, with lifetimes and energies that can be controlled by tuning the laser wavelength and intensity. Furthermore, previous studies have employed collinear and noncollinear~\cite{Leone2016TowardFWM} FWM schemes to investigate electron wave packet dynamics~\cite{materny2000wave,LeoneFWM2018H2,LeoneFWM2018N2}, molecular vibrations~\cite{bencivenga2015FWM}, coherences~\cite{LeoneNRFWM2016}, and the lifetimes of AISs~\cite{LeoneNCFWM2019}. FWM can also be used to generate coherent XUV light at new frequencies for applications.

In the conventional attosecond FWM spectroscopy the near-infrared (NIR) and XUV photons are commensurate \mbox{($\omega_{\mathrm{XUV}}\,{\approx}\,n\,\omega_{\mathrm{NIR}}$)} as the HHG spectrum consists of frequencies which are odd multiples of the driving NIR field frequency. This makes it difficult to isolate FWM emissions from the input XUV spectrum, unless noncollinear schemes~\mbox{\cite{Leone2016TowardFWM,LeoneFWM2018H2,LeoneFWM2018N2}} are invoked. In collinear geometry, we can overcome this limitation by employing an optical parametric amplifier (OPA) to produce tunable non-commensurate infrared (IR) pulses, which lead to background-free FWM emissions~\mbox{\cite{Harkema:18,tran2018phase,HarkemaPlunkett:19}}. Furthermore, the frequency tunability of the IR beam (\wir) makes it perfectly suited for controlling electronic couplings and detunings with the intermediate states~\mbox{\cite{Harkema2021,Yanez-Pagans2022}}, providing a direct knob to control FWM emission pathways.

This study uses tunable FWM spectroscopy to explore quantum beats that arise from the coherent XUV excitation of autoionizing states in krypton. The parametric conditions of the FWM processes allow us to identify the atypical composition of the autoionizing electronic wave packet (EWP) in terms of  singly-excited (SE) $4s^{-1}np$ and the doubly-excited (DE) $4p^{4}nsnp$ states. Our results expand on the previously reported photoelectron studies  in this energy regime~\cite{LeoneVMI2011,Geisler2012}. Importantly, we are able to assign the electronic configuration of doubly excited states discussed  in~\cite{CodlingMadden1971,CodlingMadden1972}. The tunability of the IR allows us to explore the role of intermediate dark states, and the relative amplitudes of single and multi-electron transitions. Our findings provide new avenues to study and control autoionizing EWPs.

The paper is structured as follows. Section~\ref{sec:exp_setup} describes the tunable FWM experimental setup. Sec.~\ref{sec:ses_and_des} discusses the states of Kr that play a key role in the observed phenomena, specifically a series of autoionizing SE and DE states and review the challenging features found in the observed signal. Sec.~\ref{sec:theory} discusses a minimal semi-empirical method, based on multichannel quantum defect theory (MQDT), to describe the dynamics of wave packet as it interacts with the two laser pulses. In Sec.~\ref{sec:results} we highlight the main features observed in the experimental signal, discuss the theoretical model that explains these features. We show a good agreement between the prediction and observed signals. The conclusions in Sec.~\ref{sec:conclusions} summarize our findings and discuss possible future applications.


\section{Experimental setup}\label{sec:exp_setup}

\begin{figure}[t]
\centering 
\includegraphics[width=\columnwidth]{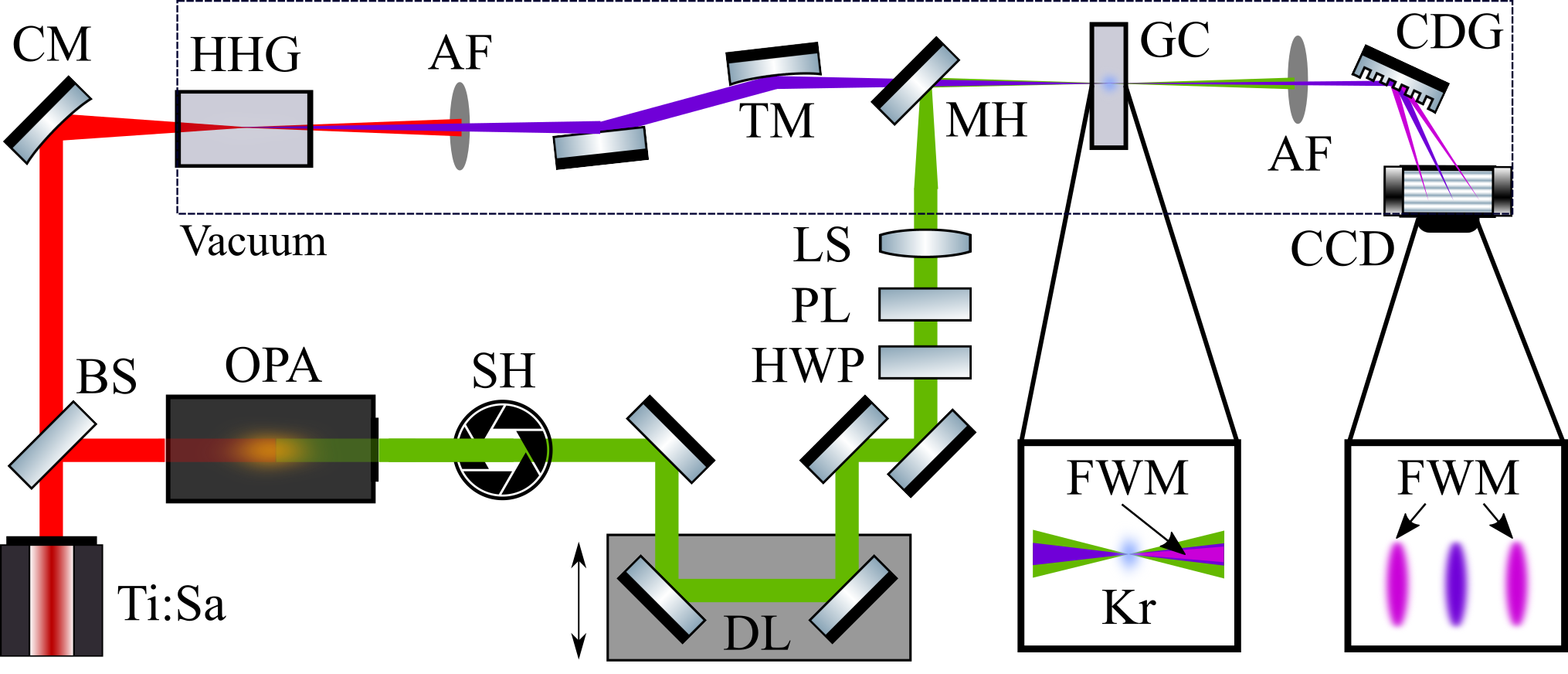}
    \caption{Experimental setup for tunable FWM spectroscopy. The fundamental NIR pulse is split in two parts. One arm (red) is used to generate XUV APT (purple) through HHG. In second arm, it is converted into tunable IR pulses (green) using an OPA. Subsequently, these pulses drive and control FWM processes in krypton, which are monitored by the XUV spectrometer.}
    \label{fig:exp_setup}
\end{figure}

Figure~\ref{fig:exp_setup} shows our tunable FWM setup, previously described in detail  in~\cite{Yanez-Pagans2022}. Briefly, we use 1.8-mJ, 40-fs, near-infrared (NIR) driving pulses with a central frequency of 1.58~eV \mbox{(785~nm)} at 1-kHz repetition rate. A 50-50 beam splitter (BS) divides the pulsed beam into two arms. One beam is focused into a xenon-filled gas cell by a curved mirror (CM) in order to produce XUV attosecond pulse trains (APT) through high-harmonic generation (HHG). The XUV phase-matching conditions maximize the photon flux of the 17th harmonic around 26.90~eV. The second  beam is passed through an optical parametric amplifier (OPA), capable of converting the NIR pulse frequency into tunable infrared (IR) frequencies in the 0.75--1.03 ($\pm$0.01)~eV range. 

Afterward, the tunable IR pulses are routed to a delay line (DL) in order to control their timing with respect to the XUV pulses. Moreover, the beam line includes a half-wave plate (HWP) and a polarizer (PL) that can be used to control the IR intensity. Subsequently, the XUV and IR pulses are collinearly combined by an annular mirror (AM) and focused into a 3-mm-thick krypton-filled gas cell by a toroidal mirror (TM) and a lens (LS), respectively. The backing pressure in the cell is \mbox{(${\sim}4$~Torr)}. A shutter (SH) in the IR path is used to alternate between XUV-only and XUV+IR cases. After the gas cell, the IR beam is filtered out by a 200-nm-thick aluminum filter (AF) and XUV continues on to the spectrometer. A concave diffraction grating (CDG) and a CCD camera are employed in the spectrometer to record the XUV spectra. 
\begin{figure}[t]
\centering 
\includegraphics[width=\columnwidth]{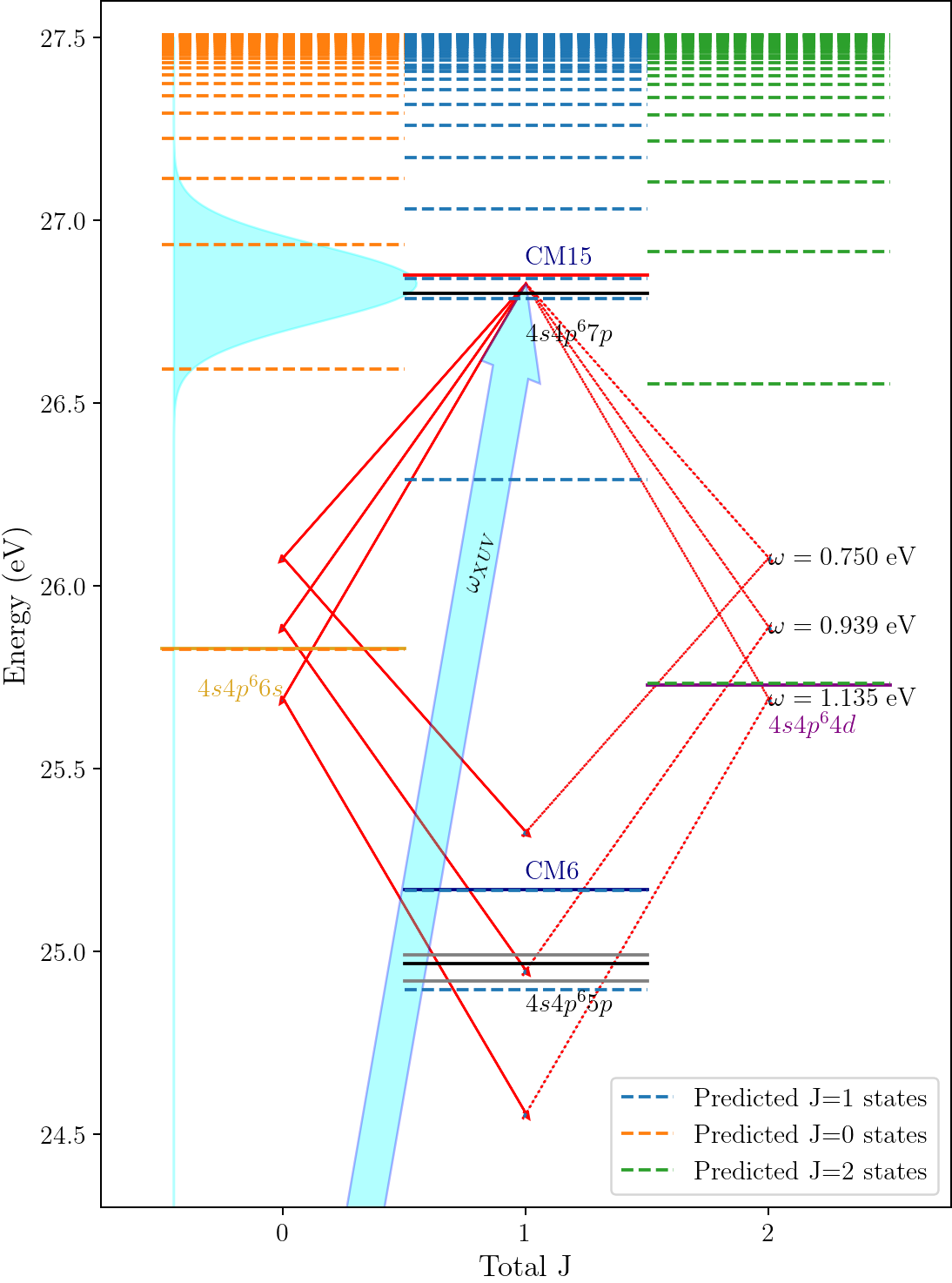}
    \caption{Energy diagram for the four wave mixing process. Solid lines indicate the experimental value of the states, and the dashed lines indicate the position predicted by the MQDT model. The arrows indicate the induced couplings by the reference 17\tsp{th} Harmonic (cyan arrow and bell curve) which creates the initial wave packet. From here the pathways induced by the IR (red arrows) interfere in the final $J=1$ states from which emission is measured in the experiment.}
    \label{fig:energy_diagram}
\end{figure}
The results are presented in terms of the normalized IR-induced changes to the XUV spectrum ($\Delta S(\omega)$), defined as $\Delta S\,{=}\,{S}_{\mathrm{XUV}}\,{-}\,{S}_{\mathrm{XUV+IR}}$. Here, both spectra are measured through the krypton-filled gas cell and normalized to $\max(|S_{\mathrm{XUV}}\,{-}\,S_{\mathrm{XUV+IR}}|)$. Positive values \mbox{($\Delta S\,{>}\,0$)} indicate IR-induced XUV absorption, negative values \mbox{($\Delta S\,{<}\,0$)} designate IR-induced XUV transparency or emission. In our scheme, the FWM emission occurs in the regions where the input XUV flux is zero~($S_{\mathrm{XUV}}\,{=}\,0$). In our analysis, negative (positive) time-delays ($\tau$) mean the IR field preceeds (succeeds) the XUV APT.


\section{Singly- and doubly-excited\linebreak autoionizing states in krypton}\label{sec:ses_and_des}

\begin{table}[b]
\caption{\label{tab:SES}SE \sestates{} states in Kr \cite{CodlingMadden1972,Sugar1991,Saloman2007}. The error is $\pm$0.02~eV.}
\begin{ruledtabular} 
\begin{tabular}{ccc}
State & Energy (eV) & Code \\
\midrule
$5p(1/2,1/2)$ & 24.92 & CM3 \\
$5p(1/2,3/2)$ & 24.99 & CM5 \\
$6p(1/2,1/2)$ & 26.30 & CM10 \\
$6p(1/2,3/2)$ & 26.31 & CM11 \\
$7p(1/2,3/2)$ & 26.80 & CM14 \\
$8p(1/2,3/2)$ & 27.03 & CM18 \\
$9p(1/2,3/2)$ & 27.18 & CM22 \\
$10p(1/2,3/2)$ & 27.27 & CM24 \\
$11p(1/2,3/2)$ & 27.33 & CM26 \\
\end{tabular}
\end{ruledtabular}
\end{table}

Our experiment explores the excitation of inner-valence SE and outer-valence DE states within the bandwidths of the XUV and IR pulses in the energy region of ~24.8--27.3~eV. Various photoabsorption and photoelectron studies have explored this spectral region due to its rich structure and abundant autoionizing dynamics~\cite{CodlingMadden1972,valin1975_22-32eV,White1983,Sugar1991,Saloman2007,LeoneVMI2011, Geisler2012}. However, several spectral lines still remain unassigned. In this energy range, as was pointed out in~\cite{CodlingMadden1971,CodlingMadden1972,valin1975_22-32eV}, there are only two possible electron configurations: single and double electron excitations. The SE states correspond to a direct XUV excitation of a $4s$ inner-shell electron to the Rydberg series converging to the $4s4p^6$ ionization threshold \mbox{(27.51~eV)} \mbox{($4s^{2}4p^{6}{\xrightarrow{\omega_{\mathrm{XUV}}}}4s4p^{6}np$)}. Due to spin-orbit coupling, for total $J=1$ ---which is the angular momentum value probed by the single photon XUV absorption--- there are two non-degenerate states. From MQDT we know that the two states will be an admixture of the two $jj$ coupled configurations $\left[4s4p^6\right]_{1/2}np_{3/2}$ and $\left[4s4p^6\right]_{1/2}np_{1/2}$, but in general the lower state will have a larger component in the $(1/2,1/2)$ state, while the larger will mostly be composed of $(1/2,3/2)$. In addition to these, the DE states are simultaneous excitations of two $4p$ valence electrons \mbox{($4s^{2}4p^{6}{\xrightarrow{\omega_{\mathrm{XUV}}}}4s^{2}4p^{4}n{\ell}n'{\ell}'$)} \mbox{\cite{CodlingMadden1971,CodlingMadden1972,Trajmar1978Kr,White1983}}. These states converge to the $4s^{2}4p^{4}5s$ ($^2P$) and ($^4P$) ionic thresholds~\cite{LeoneVMI2011}. Parallel to the states that can be directly excited from the ground states, in this energy region there are there are also dark states, with total angular momentum $J=0$ and $J=2$, corresponding to the $4s^{-1}4p^{6}nd$ and $4s^{-1}4p^{6}nd$ series. Since we aim to describe a two photon transition, these states will be relevant for the description of the observed dynamics. The position of the most relevant states (taken from the static spectral data) the laser frequencies and the states predicted from the theoretical model (that we introduce later) are shown in Fig.~\ref{fig:energy_diagram} as a reference for the excitation scheme and the relevancy of the states in the process. 

\begin{figure}[t]
\centering 
\includegraphics[width=\columnwidth]{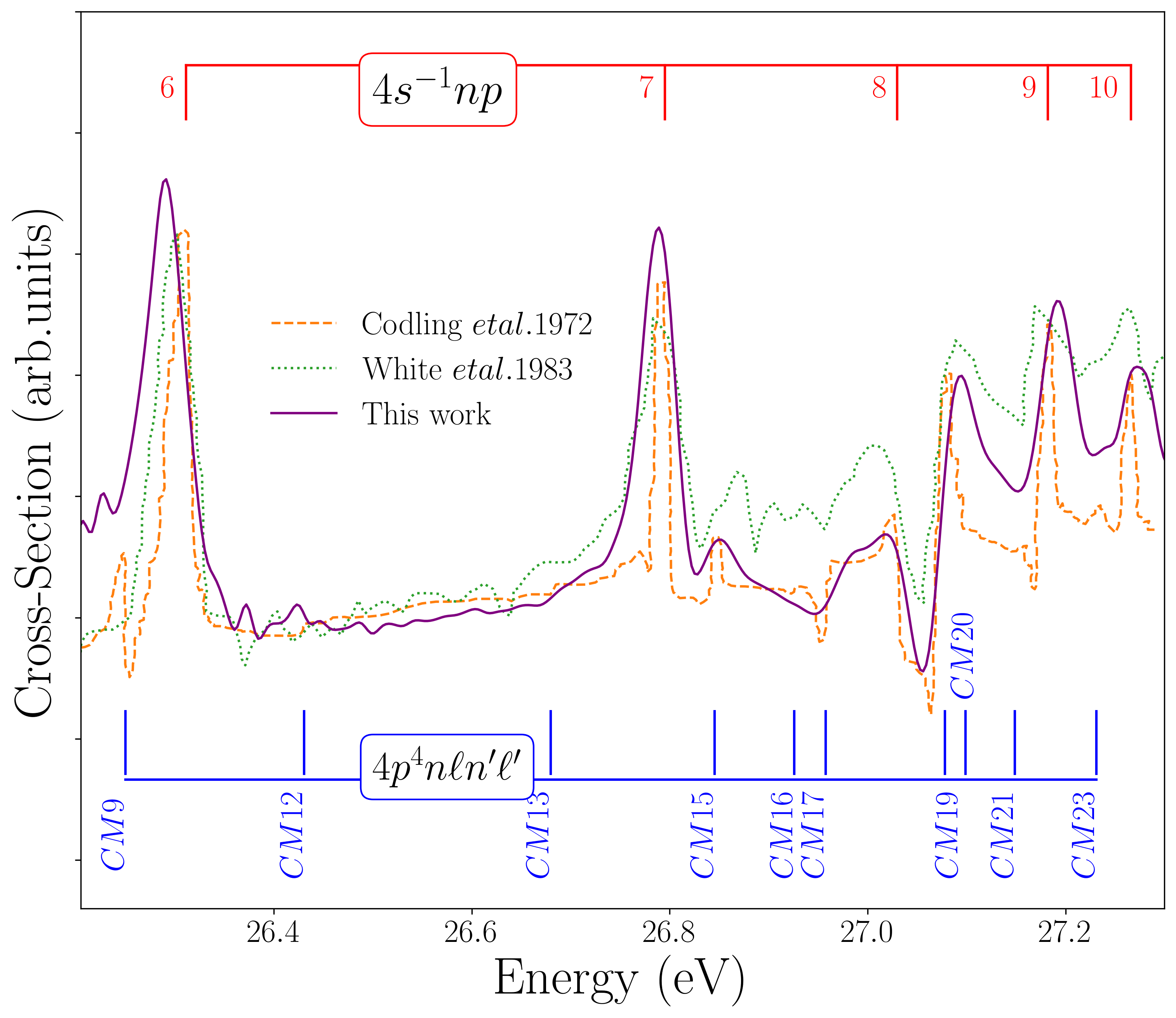}
    \caption{Reference photoabsorption and photoionization spectra of krypton in the 26.2--27.3 eV range. Our experimental photoabsorption data in vicinity of the 17th harmonic of the XUV pulse~(purple-solid line). Adapted photoabsorption from~\cite{CodlingMadden1972} (yellow-dashed line). Reference photoionization from~\cite{White1983} (green-dotted line). The features observed correspond to the \sestates{} SE (red markers) and the \destates{} DE (blue markers) states.}
    \label{fig:photoabsorption}
\end{figure}

Part of our calibration procedure involved measuring the static photoabsorption spectrum of the atom. Our measured spectrum, along with those reported in \cite{CodlingMadden1972, White1983}, is shown in Fig.~\ref{fig:photoabsorption}. We calibrated our measurement to the energies of the $4s^{-1}np$ series, finding the same enhanced absorption cross section compared to the DE states. Since the location of the additional features in the static spectrum is not substantially different from those in Codling and Madden (CM) paper\cite{CodlingMadden1972}, we will refer to them as the CM series. With this we do not imply that all the CM states belong to the same Rydberg series. The position of these states are listed in Tables.~\ref{tab:SES} and~\ref{tab:DES}.

\begin{figure*}[t]
\centering 
\includegraphics[width=\textwidth]{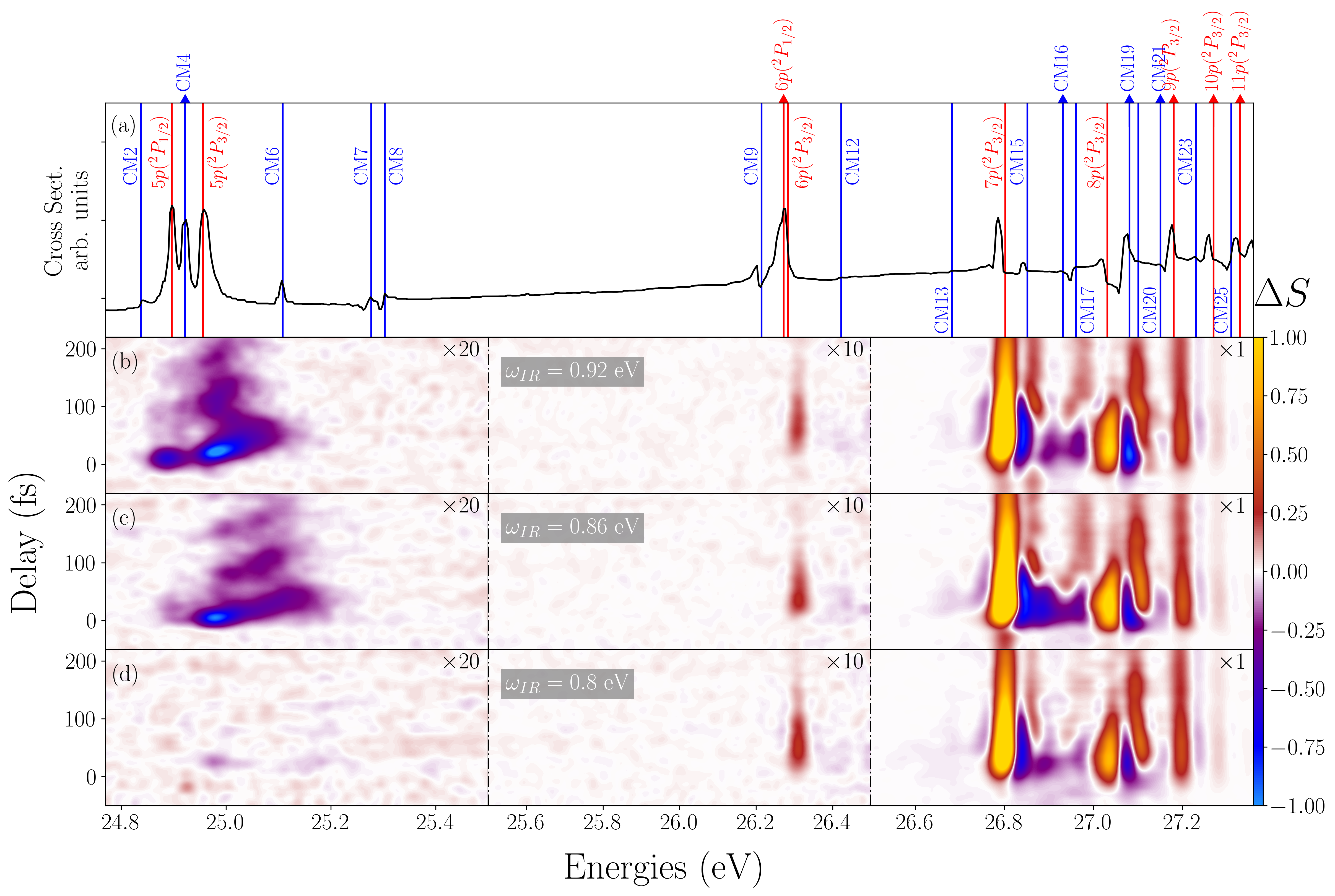}
    \caption{Tunable FWM emissions in krypton. (a) Reference static photoabsorption spectrum from \cite{CodlingMadden1972} with the CM labeled states and the $4s4p^6 np$ series. Panels (b),(c) and (d) show the time dependent difference spectrum $\Delta S$, normalized to the maximum signal value, for different IR frequencies: $1348$ nm (b), $1442$ nm (c) and $1550$ nm (d). Notice the change in strength of the signal at different energies. To bring to a similar scale the transient absorption around the $6p$ state one needs a $\times 10$ factor, while the FWM signal needed to a $\times 20$ factor.}
    \label{fig:exp_summary}
\end{figure*}

Following this calibration, we conducted time-resolved measurements. This is shown in Fig.~\ref{fig:exp_summary} where the lower three panels show the transient absorption change and FWM signals for three different IR laser frequencies, namely $1348$, $1442$ and $1550$ nm. Following the excitation from the two laser pulses, the system goes through different processes. There is transient absorption in the vicinity of the center of the XUV pulse, around $27$~eV, and the tails, near $26.3$~eV. This show some of the bright states that are reachable with the XUV pulse, but does not exhibit the strong IR dependence and rich time dynamics that the four wave mixing emission spectra, around $25$~eV, has. We focus on analyzing these features in this article. One of the most evident features, seen in panels (b)-(d) in Fig.~\ref{fig:exp_summary}, is the oscillation in time in this latter region. Other two features that warrant analyzes are the dependence of the phase with energy and the variation of the strength of the oscillations with IR frequency. To analyze it, it is crucial to understand the nature of the initial XUV excitation from the ground state and of the subsequent IR couplings. From the dipole selection rules we know that the wave packet formed from the excitation of the ground state necessarily will have $J=1$ and odd parity, therefore the relevant intermediate states will have even parity and $J=0$ or $2$. This is where the aforementioned $4s^{-1}ns$ and $4s^{-1}nd$ Rydberg series will be relevant. Thus, being able to model how these two symmetries interact with the initial wave packet will allow us to understand the source of the observed features in the spectrum. 

The complexity of the Kr atom in this energy range, prevents us from forming an illustrative \emph{ab initio} model that captures all this features and is simple enough to allow for practical understanding of the underlying mechanism. This is why we use a minimal MQDT model to understand the interactions of the active electron including the physics of the complex ionic core.
\begin{table}[b]
\caption{\label{tab:DES}DE \destates{} states in Kr \cite{CodlingMadden1972}. The error is $\pm$0.02~eV.}
\begin{ruledtabular} 
\begin{tabular}{cccc}
Code & Energy (eV) & Code & Energy (eV) \\
\midrule
CM2 & 24.85 & CM15 & 26.85\\
CM4 & 24.95 & CM16 & 26.93\\
CM6 & 25.17 & CM17 & 26.96\\
CM7 & 25.37 & CM19 & 27.08\\
CM8 & 25.40 & CM20 & 27.10\\
CM9 & 26.25 & CM21 & 27.15\\
CM12 & 26.43 & CM23 & 27.23\\
CM13 & 26.68 & CM25 & 27.31\\
\end{tabular}
\end{ruledtabular}
\end{table}
\section{Theoretical Model}\label{sec:theory}
\subsection{Wave functions and Dipole moments}
In order to capture the dynamics involved in this experiment, we propose a minimal multichannel quantum defect model (MQDT). Since the problem involves the excitation of a wave packet from the ground state and the subsequent interaction with the IR pulse, several symmetries are required. 

As we will discuss later, we choose to characterize the different symmetries by a $K$ matrix that defines the asymptotic form of the linearly independent solutions, determining the position of autoionizing resonances and their width. Thus, we will use the position of the resonances to fit the entries of the $K$ matrix on each symmetry. To more easily deal with the possible divergence of the $K$ matrix, we parametrize it with the $\mu$ matrix, see for example \cite{PRXQuantum.3.020326}, which is related with the matrix function $K=\tan \pi \mu$.  

The first symmetry describes the wave packet excited from the ground state by the XUV. It has odd parity and $J=1$. The parameters are fitted to the position and width of the resonances $4s^{-1}7p$, $4s^{-1}5p$ and $CM15$ as presented in Fig.~\ref{fig:photoabsorption}. The three channels used to describe this symmetry are shown in Table~\ref{tab:params1}, approximating the continuum with a single open channel. The two closed channels are assumed to correspond to the $4s^{-1} 4p^6 np$, and to the $4s^2 4p^4 5s np$ electronic configurations. The parameters used to fit the position and width of the resonances are shown in Table~\ref{tab:params1}. In order to fit both the $5p$ and $7p$ states the parameters must be energy dependent, and it shows evidence that state CM4 belongs to the same Rydberg series as CM15.

\begin{table}[]
    \centering
    \begin{ruledtabular}
    \begin{tabular}{cc c c}
    & $4s 4p^6 np$ & $4s^24p^4 5s np$ & $4s^2 4p^5 \epsilon s/d$ \\ \midrule 
         $I$ & $27.5139$ & $28.1807$ & $14.2215$ \\\midrule
         $\mu^{(0)}_{i j}$ & -0.150 & 0.243 & 0.375  \\
         & 0.243 & 0.225 & 0.461 \\
         & 0.375 & 0.461 & 0.164 \\[4mm]
         $\mu^{(1)}_{i j}$ & 0.0279 & -0.055 & 0  \\
         & -0.055 & -0.130 & 0 \\
         & 0 & 0 & 0 \\
    \end{tabular}
    \end{ruledtabular}
    
    \caption{Energy dependent quantum defect ($\mu$-) matrix with channel labels and their corresponding ionization thresholds ($I$, in eV) for the $J=1$ odd symmetry. The threshold for the second channel is obtained by a weighted average over the different $J$ values. The energy dependence is linear and centered at the $CM15$ energy  (26.8400 eV). The derivative term $\mu^{(1)}$ is given in eV$^{-1}$.} 
    
    \label{tab:params1}
\end{table}

The two additional symmetries have even parity and, $J=0$ and $2$. They will describe the intermediate states that mediate in the two photon process that we want to describe. Experimental evidence from electron collision experiments \cite{CEBrion_1970} indicate that for a complete description the $J=0$ symmetry must at least include the closed $4s4p^6 ns$ channel and the $J=2$ symmetry must include $4s4p^6 nd$. In principle a doubly excited channel should also be included, but the only possibility for an outer electron to be coupled with the $4s^2 4p^4 5s$ ionic core to produce the correct symmetry are $s$ and $d$, nonetheless in the energy region the effective quantum number relative to this threshold is around $2.5$ which makes it unlikely for a state with such angular momentum to be attached.\\

Consequently, the intermediate channels are described using two channel models and the parameters are fitted to the positions and widths of the intermediate $4s4p^6 6s$ and the $4s4p^6 4d$ state. Since these are the only states that could be in close resonance with the IR pulse the $K$ matrix is taken to be constant. The value of these parameters is given in Tab.~\ref{tab:params2_2} and Tab.~\ref{tab:params2_0}. \\

Briefly, we mention that the numerical fits for the $J=1$ odd parity symmetry was conducted by determining the position of the resonances by directly solving the generalized eigenvalue problem detailed below. For the two channel models, it is possible to find analytical expressions for the position of the resonances that we fit to the data. In the appendix~\ref{app:B} we detail two different analytical forms that give the quantum defect as a function of the energy independent parameters.\\

\begin{table}
    $$J=2$$
    \begin{ruledtabular}
        \begin{tabular}{ccc}
         & $4s 4p^6 nd$ \quad \quad & \quad \quad $4s^2 4p^5 \epsilon p/f$ \\ \midrule
         $I$ & $27.5139$ &$14.2215$ \\ \midrule
         $\mu^{(0)}_{i j}$ & -0.262 & 0.382  \\
           & 0.382 & -0.262\\
    \end{tabular}
    \end{ruledtabular}
    
    \caption{Channels and energy independent parameters found to fit the $4d$ $J=2$ resonance.}
    \label{tab:params2_2}
\end{table}

\begin{table}[]
    \centering
    $$J=0$$
    \begin{ruledtabular}
        \begin{tabular}{ccc}
         & $4s 4p^6 ns$ \quad \quad & \quad \quad $4s^2 4p^5 \epsilon p/f$ \\ \midrule
         $I$ & $27.5139$ &$14.2215$ \\\midrule
         $\mu^{(0)}_{i j}$ & -0.825 & -0.168  \\
           & -0.1678 & -0.770 \\
    \end{tabular}
    \end{ruledtabular}
    
    \caption{Channels and energy independent parameters found to fit the $6s$ $J=0$ resonance.}
    \label{tab:params2_0}
\end{table}
We will model the observed signal, by calculating the two photon transition amplitude following the excitation of the XUV. Since this whole process occurs above the $4s^24p^5$ threshold, the appropriate boundary condition to impose on all the wave functions is the the incoming wave boundary condition. The procedure to impose this boundary condition given a $K$ matrix is standard in MQDT literature, and we briefly outline it here.\\

Before imposing boundary conditions at infinity, for any energy $E$ there exist as many solutions as there are channels in the symmetry. There is freedom to choose any linearly-independent functions to expand the physical solution in terms of, and we choose the basis of real, \emph{standing wave $K$-matrix solutions} represented in terms of regular and irregular Coulomb functions $(f,g)$ in each channel at electron distances outside the ion, at $r>r_0$ as follows:
\begin{equation}
    \psi^{K}_i(E,r) \underset{r\to\infty}{\to} {\mathcal{A}}\sum_{j=1}^{3} 
    \frac{\Phi_{j}}{r} \left[f(\epsilon_j,\ell_j,r) \delta_{ij} -  g(\epsilon_j,\ell_j,r)K_{ji} \right], 
\end{equation}
or in a standard matrix notation (see, e.g. \cite{Aymar1996}) as 
\begin{equation}\label{eq:nonphys}
    \psi^{K}= {\mathcal{A}} {\bf \Phi} ({\bf f}-{\bf g} {\bf K}).
\end{equation}
Here ${\bf f}$ and ${\bf g}$ are diagonal matrices with the regular and irregular Coulomb functions, respectively, and ${\mathcal{A}}$ is the antisymmetrization operator. Further, $\epsilon_j=E-I_j$ is the Rydberg electron energy and $\ell_j$ is the electronic angular momentum in the $j-$th channel represented by the channel function $\Phi_j$, which encompasses all angular as well as ionic degrees of freedom.\\


The $K$-matrix solutions do not obey physical boundary conditions, since the radial component of the closed channels (i.e. $\epsilon_i < 0$) diverges; thence, we need to remove this divergence. This is achieved by taking linear superpositions of the un-physical solutions to eliminate the diverging terms. In the end there will be as many solutions as there are open channels, call it $N_o$, and they have an asymptotic form that can be expressed as the formal solutions in Eq.~\ref{eq:nonphys}
\begin{equation}
    \psi^{K,{\rm phys}} \underset{r\to\infty}{\to}  {\mathcal{A}} {\bf \Phi} ({\bf f}_o-{\bf g}_o {\bf K}^{\rm phys}(E))
\end{equation}
a key difference is that the matrix $K^{\rm phys}$ depends strongly with energy, since it includes all the Rydberg series information. It is  expressed in terms of the formal $K$ matrix by a standard MQDT formula
\begin{equation}\label{eq:kphys}
    K^{\rm phys}(E)=K_{oo}-K_{oc}(K_{cc}+\tan{\beta_c(E)})^{-1}K_{co}
\end{equation}
where the sub-indices in the $K$ matrix indicates whether it correspond to open-open (oo), closed-closed (cc), open-closed (oc) or closed-open (co) coupling. The closed channels are defined by their close range phase shift $\beta_i \equiv \pi (\nu_i -\ell_i)$, where the effective quantum number is given by $\nu_i \equiv (-2\epsilon_i)^{-1/2}$. \\

Explicitly, the linear superposition that gives the $N_o$ solutions $\psi^{K,{\rm phys}}$ can be obtained by right-multiplying the original $N$ $K$-matrix solutions in Eq.\ref{eq:nonphys} by an $N \times N_o$ matrix that equals the $N_o \times N_o$ identity matrix in the first $N_o$ rows and which in the final $N_c$ rows is equal to  the matrix $-(K_{cc}+\tan{\beta_c})^{-1}K_{co}$. This transformation will also be needed to get the correct final electric dipole matrix elements.\\

In some of the formulas below it is convenient to introduce the energy-orthonormal collision eigenchannel solutions, which are expressed in terms of the normalized eigenvectors $T_{j\rho}$ and eigenvalues $\tan{\pi \tau_\rho}$ of $ K^{\rm phys}(E)$, namely:
\begin{equation}\label{eq:eigc}
    \psi^{\rm eig}_\rho(E) = \sum_j \psi^{K,{\rm phys}}_{j} T_{j\rho} \cos{\pi \tau_\rho},
\end{equation}
or more compactly, $\psi^{\rm eig}(E) = \psi^{K,{\rm phys}} {\bf T}  \cos{\pi {\bf \tau}}$. Furthermore, the incoming wave boundary condition wave function which represents an electron leaving the core in a specific channel $\Phi_j$ can be expressed in therms of these eigen functions by
\begin{equation}
    \psi_j^{-}(E) = \sum_\rho e^{-i\sigma_j-i\pi \tau_\rho} T_{j\rho}\psi^{eig}_\rho(E),
\end{equation}
these functions will be the correct solution to describe the final states that can reach the detector at specific direction for the photo-electron. On the other hand the initial state will have a direction specified by the exciting laser, and so the outgoing wave boundary condition state 
\begin{equation}
    \psi_j^{+}(E) = \sum_\rho e^{i\sigma_j+i\pi \tau_\rho} T_{j\rho}\psi^{eig}_\rho(E),
\end{equation}
will be the appropriate physical state. 

For the intermediate states, since we will be integrating over them, the eigen channel wave physical solution in Eq.~\ref{eq:eigc} will be sufficient. 

For the case that pertains to us, which consists of a single open channel, the eigen channel solutions and the physical incoming wave boundary condition solutions are related by a complex scaling, so we will focus on expressing the dipoles in terms of the eigen channel functions.\\

For this case the physical K matrix of equation \ref{eq:kphys} will be a scalar, not a matrix, from which the eigen phase can be extracted by $\tau(E) = 1/\pi \arctan K^{\rm phys}(E)$. Assuming the last channel is the open channel, we can write explicitly Eq.~\ref{eq:eigc} as 
\begin{equation}
\begin{aligned}
    \psi^{eig}_\rho &= -\cos \pi \tau \sum_{i=1}^{N-1} \psi_i^{K}\left(\frac{K_{co}}{K_{cc}+\tan \beta}\right)_i  \\
    &+ \psi_N^{K},
\end{aligned}
\end{equation}
which we can compactly write as $\psi^{\rm eig}_{\rho} = \{\psi^{K}_i\}\cdot \vec{X}$. If we have access to the reduced dipole moments between different primitive $K$ matrix states, the dipoles we are interested can be calculated by using the the matrix transformation. For the initial state with an intermediate state we have 
\begin{widetext}
\begin{equation}
    \left\langle\psi^{eig,J'}(E')\middle|\middle|D\middle|\middle|\psi^{+,J}(E)\right\rangle = e^{i\sigma(E)+i\pi \tau(E)} \sum_{i,j} X^{J'}_i \left\langle\psi^{K,J'}_i\middle|\middle|D\middle|\middle|\psi^{K,J}_j\right\rangle X_j^{J},
\end{equation}
and similarly for the final state we have 
\begin{equation}
    \left\langle\psi^{-,J'}(E')\middle|\middle|D\middle|\middle|\psi^{eig,J}(E)\right\rangle = e^{i\sigma(E')+i\pi \tau(E')} \sum_{i,j} X^{J'}_i \left\langle\psi^{K,J'}_i\middle|\middle|D\middle|\middle|\psi^{K,J}_j\right\rangle X_j^{J}.
\end{equation}
The dipole matrix element can be calculated using these reduced matrix elements by using the Wigner-Eckart. Putting it all together, gives an expression for the dipoles that will define the two photon transition 
\begin{equation}
\begin{aligned}
        \bra{\psi^{-(J')}(E')}z\ket{\psi^{\rm eig,(J)}(E)} = (-1)^{J'-M'} e ^{i\sigma(E')+i\pi \tau(E')}\begin{pmatrix} J' & 1 & J \\ -M' & 0 & M\end{pmatrix} \left(X^{J'} \middle|\middle| D\middle|\middle|X^J\right), \\
        \bra{\psi^{\rm eig,(J')}(E')}z\ket{\psi^{+,(J)}(E)} = (-1)^{J'-M'} e ^{i\sigma(E)+i\pi \tau(E)}\begin{pmatrix} J' & 1 & J \\ -M' & 0 & M\end{pmatrix} \left(X^{J'} \middle|\middle| D\middle|\middle|X^J\right), \\
    \end{aligned}
\end{equation} 
\end{widetext}
where matrix $D$ is the matrix of reduced matrix dipole elements between the standing wave $K$-matrix solutions. The inner product with the round parenthesis denote the regular matrix inner product of vectors, and we keep the double line notation of the reduced matrix elements avoid confusions.  We recognize that this decomposition of the reduced dipole matrix could lead to ill definitions, since it involves dipoles between closed channels that in principle diverge, but it's a clear decomposition of a converging quantity that allows for the identification of relevant couplings to analyze the phenomenon observed in the experiment. 

\subsection{Definition of the wave-packets}
We will treat the interaction with both the pump and the probe pulses perturbatively. Assuming that the Hamiltonian of the system is given by
\begin{equation}
\begin{aligned}
    H = H_{atom} + \mathcal{F}_{UV} \exp\left[-\left(\frac{t}{\gamma_{UV}}\right)^2\right] \cos \omega_{UV} t\\ + \mathcal{F}_{o} \exp\left[-\left(\frac{t-t_o}{\gamma}\right)^2\right] \cos \omega (t-t_o),
\end{aligned}
\end{equation}
one can show that if the initial state of the system at $t=-\infty$ is the Kr ground-state, then the wave packet excited for $0<t<t_o$ is given by
\begin{equation}
    \psi_o = \int c(\delta) \psi^{-,(1)}(\delta) \text{  } d\delta
\end{equation}
where the amplitudes are given by
\begin{widetext}
    \begin{equation}
     c(\delta)= \sqrt{\frac{\pi}{3}} \frac{1}{T(\delta)} \mathcal{F}_{UV}  \exp \left[-\left(\gamma_{UV}\frac{\delta-\omega_{UV} }{4}\right)^2 + i\sigma(\delta)+i\tau(\delta)\right]\\ \sum_{j=1}^3 \left\langle\psi^{(1)}_i\middle|\middle|z\middle|\middle|\psi_o\right\rangle C^{(1)}_{i}(\delta)
\end{equation}

To account for the effects of the probe we have to calculate the amplitude in the continuum of the $J=1$ symmetry for times long after the probe laser has passed. For this, we need to  use second order time dependent perturbation theory. The amplitude is given by the known time ordered integral expression \cite{bello2022electronic}, which upon replacing the the dipole matrix elements we derived in the previous section gives


\begin{equation}\label{eq:TPA}
\begin{aligned}
        T(E,t_o) = 2 \pi \mathcal{F}_o^2 \gamma^2 \sum_{J=0,2}
    \begin{pmatrix}
        1 & 1 & J \\
        0 & 0 & 0
    \end{pmatrix}^2 
    \int d\delta \quad  c(\delta) \exp{-\frac{\gamma^2}{8}(E+2\omega-\delta)^2 + i (E-\delta)t_o} \\[2mm]
    \int d\chi \quad \overline{\mathcal{W}(\chi,E)} \mathcal{W}(\chi,\delta)\text{  } w\left[\frac{(E+\delta-2 \chi) \gamma}{\sqrt{8}} \right] \left(X^{(1)}_E\middle|\middle|D\middle|\middle|X^{(J)}_\chi\right) \left(X^{(J)}_\chi\middle|\middle|D\middle|\middle|X^{(1)}_\delta\right)
\end{aligned},
\end{equation}
\end{widetext}
where $\chi$ represents the integration over the intermediate energies, and $w$ is the Faddeva function \cite{NIST:DLMF}.

\begin{table}[]
    \centering
    \begin{ruledtabular}
    \begin{tabular}{ccc}
         State & Energy (eV) & Width (meV)  \\
         \midrule
         $4s4p^6 5p$ & 24.969& 177\\
         $CM6$ & 25.170 & 29\\
         $4s4p^6 6p$ & 26.274 & 49\\
         $4s4p^6 4d$ & 25.733 & 125\\
         $4s4p^6 6s$ & 25.838 & 249\\
         $4s4p^6 7p$ &26.785 & 28\\
         $CM15$ & 26.840 & 24\\
    \end{tabular}
    \end{ruledtabular}
    \caption{Theoretical estimated energy and width of the most relevant states involved in our study.}
    \label{tab:my_label}
\end{table}
Just as a reference see the horizontal dashed lines in Fig.~\ref{fig:energy_diagram} that show the predicted position of the resonances. One sees that the parameters obtained also account for the position of the $4s 4p^6 6p$ state, and that higher members of the $4s 4p^6 nd$ and $4s 4p^6 ns$ series are not as well approximated due to the energy dependence of their quantum defects.


\section{Results and Discussion}\label{sec:results}

Revisiting Figure~\ref{fig:energy_diagram}, we see the energy diagram for tunable coupling pathways that lead to background-free FWM emissions in krypton. Initially, the XUV pulse (cyan spectrum and arrow), centered around 26.90~eV, coherently excites several singly and doubly excited states in the 26.2--27.3~eV spectral region. These correspond to the members of the \sestates{} ($n\sim 7-10$) and what we propose are doubly excited states with dominant electronic configuration $4p^45snp$ ($n\sim 6-7$). Following their coherent excitation, a tunable time-delayed IR pulse (different red arrows) couples these states to the two series of intermediate dark states (orange and green dashed lines) with total $J$ equal to $2$ or $0$. The IR pulse is long and strong enough to allow for a two-photon process resulting in population transfer to lower members of the SE and DE series, more specifically $4s^{-1}5p$ and $4p^45s5p$. From here, these AIS can radiate back to the ground state resulting in the measured background-free FWM radiation measured in this experiment. The measured emission spectrum is shown in Fig.~\ref{fig:exp_summary} where signal oscillates with delay, has a phase that is dependent with the photon energy and the amplitude changes with the frequency of the IR pulse. \\

\begin{figure}
\centering 
\includegraphics[width=\columnwidth]{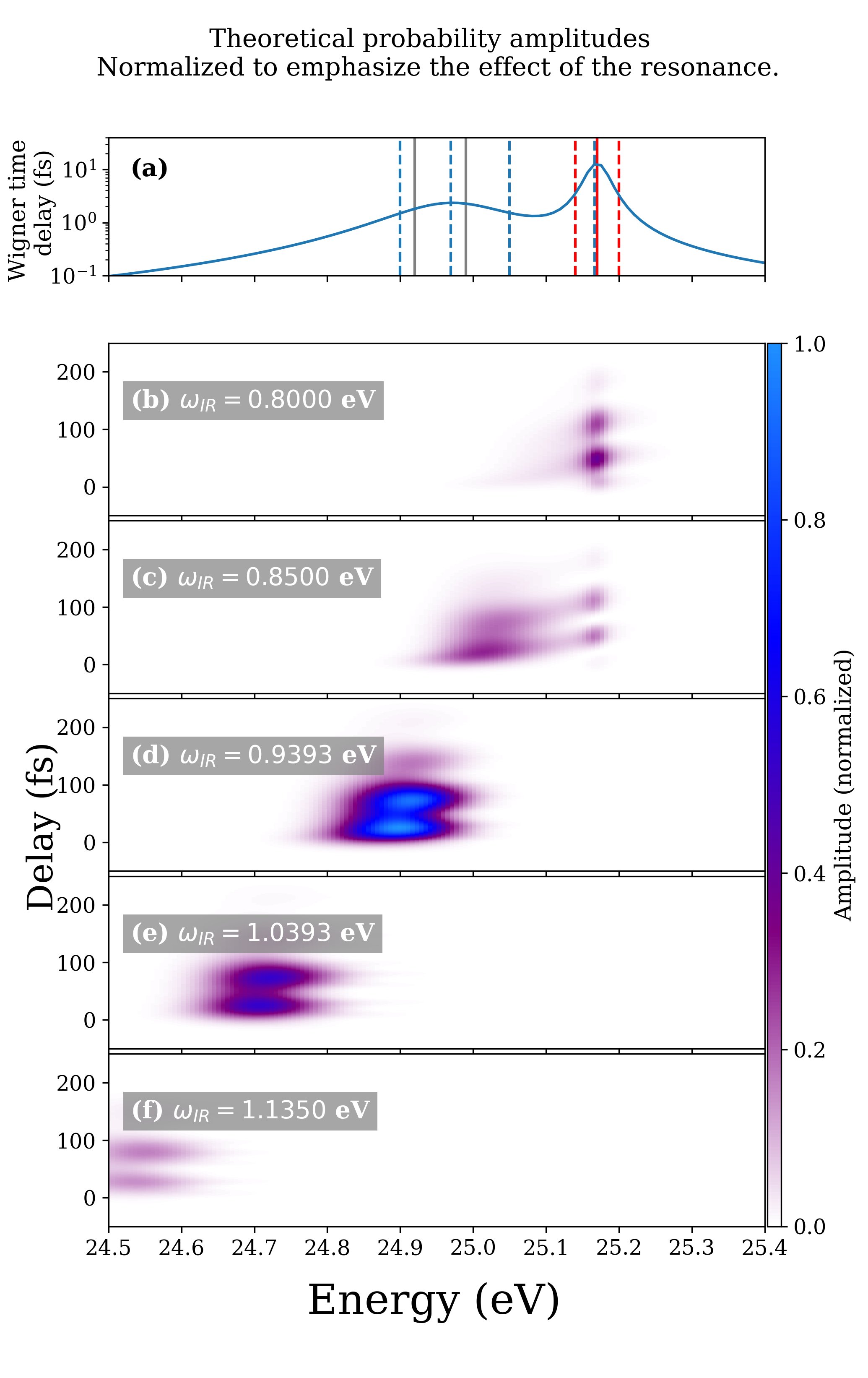}
    \caption{Theoretical calculation of the population of states at different energies for different values for the IR frequencies. On top we show the time delay calculation and vertical lines indicating the position of the CM15(Red) and the $5p$ states (grey lines). The subsequent panels (b-f) show different frequencies of the IR and the produced oscillations.}
    \label{fig:theory}
\end{figure}

Panel (a) of Fig.~\ref{fig:exp_summary} shows a reference static photo absorption spectrum from \cite{CodlingMadden1972}, with vertical lines indicating the position of the known resonances. Panels (b)-(d) show the resulting delay-dependent measurement of the FWM signal, expressed as a difference between the response of the system with the pump and the probe and that only with the pump, for three different frequencies of the IR pulse. Panel (b) corresponds to $1348$~nm~($0.92$~eV), panel (c) shows $1442$~nm~($0.86$~eV) and panel (d) contains $1550$~nm~($0.8$~eV). The oscillation period is seen to correspond to the energy separation between the the CM15 feature and the $4s^{-1}7p$ resonance, providing evidence of their excitation through the XUV pulse and the existence of a pathway that leads to quantum interference between doubly and singly excited states. Another notable feature is that as the IR transition from the \sesp~or CM15, to the dark \sess~state is brought within bandwidth of the IR pulse the intensity of the emission is enhanced. This provide evidence to the fact that the IR mediated pathways lead to interference observed in the FWM emission and that the strength and structure of the signal will depend on the intermediate resonances. Finally, one can see that as the pulse probes through the $5p$ resonances, we observe that the oscillations posses a phase that depends on energy, to the point that the phase between the center of the signal at the energy of the CM6 feature is almost exactly out of phase from the $5p$ state.\\ 

The theoretical model aims at describing this two photon process by modeling the single atomic response to the sequence of pulses. We approach this by determining the initial amplitude of XUV excitation through a first order perturbative calculation, and then calculating the two photon amplitude of the IR process, following the derivation in section~\ref{sec:theory}. We find that the population for the energies around the CM6 and the $5p$ states oscillates with time delay at frequencies corresponding to the energies that are coherently excited by the XUV pulse. This can be observed by taking the norm squared of the transition amplitude, Eq.~\ref{eq:TPA}, where the $e^{iEt_o}$ term cancels and the oscillatory term $e^{i(\delta-\delta')t_o}$ comes out explicitly. Performing this calculation directly is not convenient, given that time delays considered in the experiments makes for very fast oscillations that very quickly make the integration over $\delta$ numerically untractable. For the present study, we use the fast Fourier transform algorithm to calculate Eq.~\ref{eq:TPA} and then take the norm squared of this quantity. This restricts our time delay resolution to the choice of energy mesh, but it was chosen as to be comparable with the experimental resolution. The resulting spectrogram is shown in Fig.~\ref{fig:theory} where each of the panels (b)-(d) shows the result for a different IR frequency, similar to experiment, demonstrating that not only does the intensity of the population changes with frequency but also the structure of the oscillation regarding the phase dependence of the energy. Indeed, the parameters we determined fit the static spectrum and the dipoles that were found to describe the coupling between the different symmetries captures this energy dependent phase in the oscillation and the dependence of the amplitude with IR frequency. We identify the reason for this as a combination of the relative signs between the reduced dipole elements, shown in Tab.~\ref{tab:dipoles}, for the different channels and the intrinsic phase (present in the $e^{i\sigma+i\tau}$ term in the formulas above). The most relevant explanatory point comes from noticing that in order to have a phase difference, the dipoles between the $J=0$ and $J=1$ each closed channels have the same relative sign, while those between $J=2$ and $J=1$ have opposite signs. \\

The width of the single $5p$ resonance in the theory is much larger than in the experiment, and thus we find a wider range of IR frequencies that show a strong signal. This effect could be modified once the spin orbit splitting is included in the calculation. This is left for future investigation. We must also notice that calculated signal is dependent on the value of the reduced dipole matrix elements. The values found here, shown in Tab.~\ref{tab:dipoles}, were obtained by performing a random search and comparing with the measured signal. Therefore, one should read this parameters as very rough estimates for these values, but they provide starting points for future investigations and as we discussed above give some pointers to explain the relation between this parameters and the features of the experimental spectrum.

\begin{table}[]
    \centering
    \begin{ruledtabular}
    \begin{tabular}{c c c}
     \diagbox[]{J=1}{J=0} & 1 & 2   \\
     \midrule
     1 & 1.492 & -1.234 \\
     2 & -0.361& 0.248\\
     3 & 0.195 & 0.0695 \\
    \end{tabular}
    \end{ruledtabular}
    \\[5mm]

    \begin{ruledtabular}
    \begin{tabular}{c c c}
     \diagbox[]{J=1}{J=2} & 1 & 2   \\
     \midrule
     1 & -2.73 & -0.271 \\
     2 & 0.955 & -1.664\\
     3 & 0.107 & 0.243 \\
    \end{tabular}
    \end{ruledtabular}
    
    \caption{Estimated values for the reduced dipole moments between the different channels for each one of the symmetries.}
    \label{tab:dipoles}
\end{table}

\begin{figure}
    \centering
    \includegraphics[width=\linewidth]{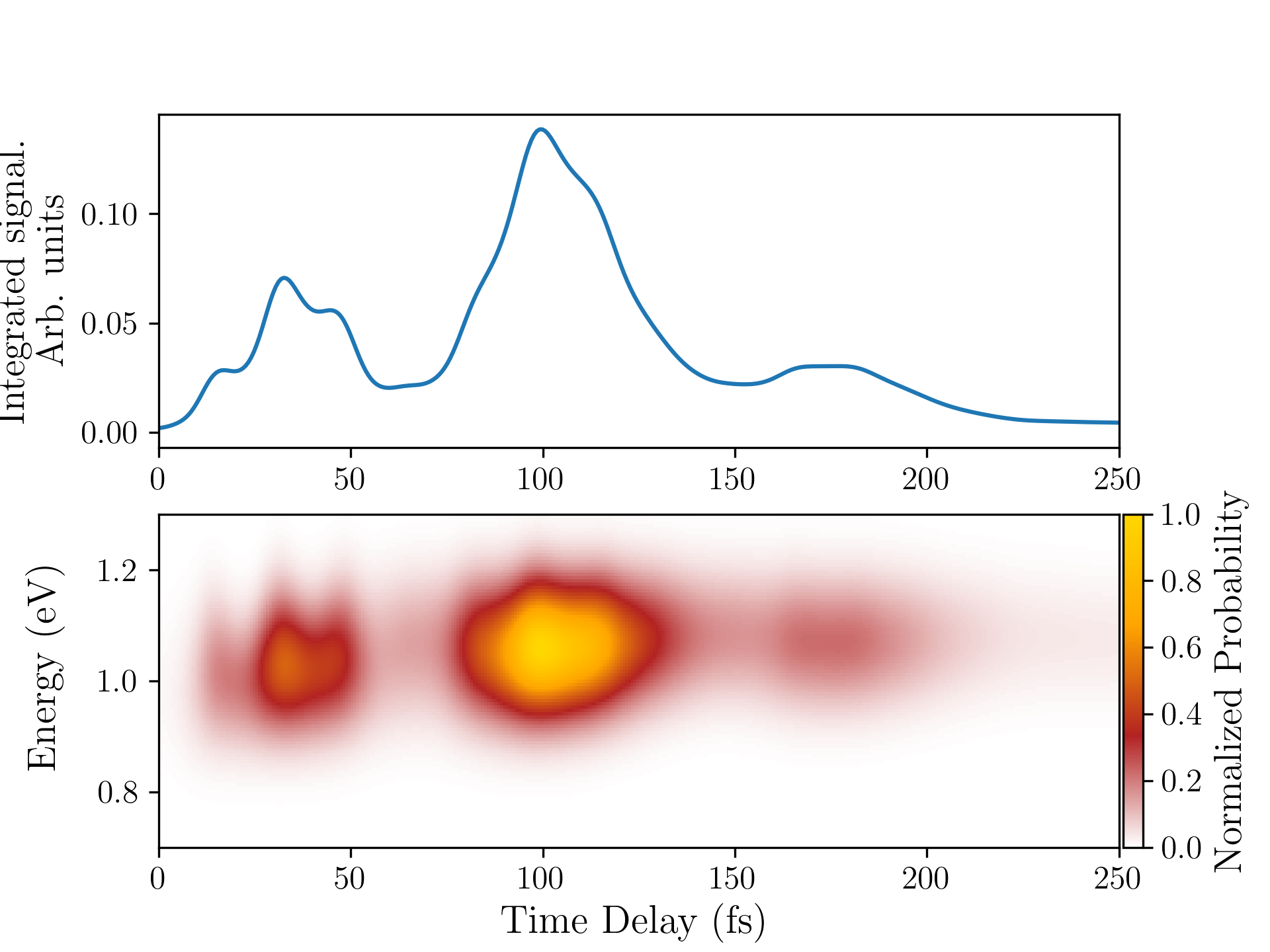}
    \caption{Calculated one photon photoionization probability for ionization from the same initial linear combination of the $J=1$ states.}
    \label{fig:photo}
\end{figure}

\begin{figure}[t]
    \centering
    \includegraphics[width=\columnwidth]{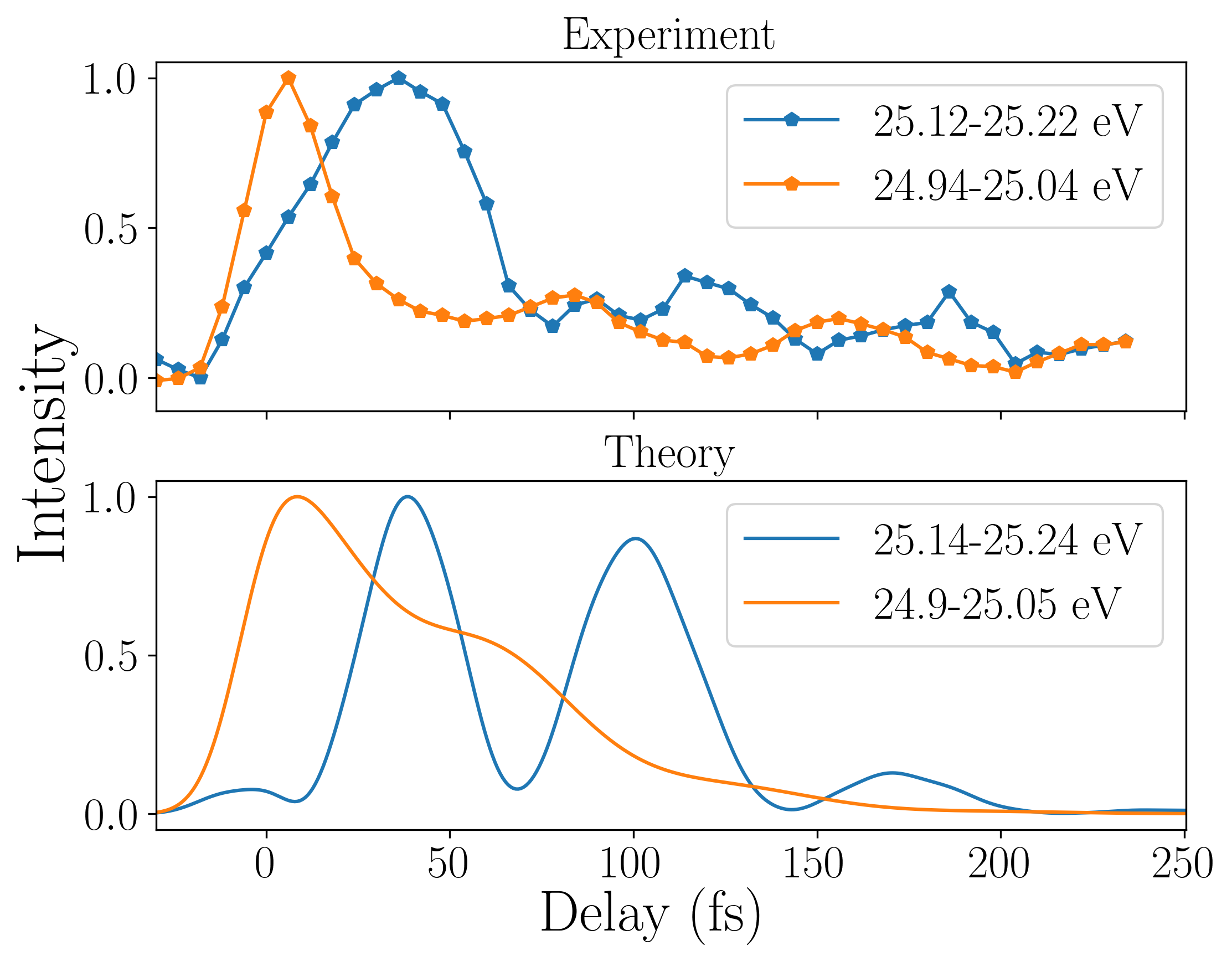}
    \caption{Line outs obtained by averaging across the energy range indicated in the caption for the experimental (top) and theoretical (bottom) data. It highlights the similar difference between the phases in the two energy regions and the decay rate which is determined by the width of the autoionizing states excited by the pulse.}
    \label{fig:lineouts}
\end{figure}

Now we can compare the quantum beats obtained in the FWM signals in theory and experiment. We show in Fig.~\ref{fig:lineouts} a comparison between the average of the signal across the width of the signals that correspond to the $5p$ states and for the CM6 in experiment and theory. For experiment, we use the signal corresponding to a $1442$~nm ($0.86$~eV) IR shown in in panel (c) of Fig.~\ref{fig:exp_summary} and use the signal corresponding to the energy of $5p(^2P_{3/2})$ state to take the lineout representing this state. For the theory we use panel (c) of Fig.~\ref{fig:theory} where a clear separation between the population on both states exists.To calculate the line outs we average across the width of the feature. The limits for this averaging are depicted by the dashed lines (blue for the $5p$ state and red for CM6) in panel (a) of Fig.~\ref{fig:theory}. In both cases we observe a phase difference between the oscillations of roughly $\pi$ and see that the oscillation peaks for delays that are larger than $0$ fs. We also observe that the decay rate of the signal is in a similar scale of around $200$~fs. In another study where single photon ionization was studied in this system \cite{Geisler2012}, a similar time scale for decay was found and was dictated by the width of the states excited by the XUV from the ground state.

In order to determine the accuracy and capabilities of our minimal model, we present estimations for the single photon ionization cross section from \cite{Geisler2012}. This study also shown similar oscillations following ionization after absorption of a single IR photon after excitation from the XUV. We leave the details of the calculations are presented in the Appendix~\ref{app:B}, and focus on the obtained photoionization probability is shown in Fig.~\ref{fig:photo}. There we show the average over the energy on the top panel. In order to parallel the results of \cite{Geisler2012}, we fit their eq.~(1) to our data and find the widths of the states to be given by $\Gamma_1=\Gamma_2=12.861$meV, $\bar{\Gamma}=14.698$meV, $a=3.89$, $b=3.189$ and $\phi=1.854$. These widths compare favorably to the widths computed by fitting a Lorentzian to the time delay curve obtained from an MQDT analysis. For the two relevant states we find $\Gamma_{7p}=24$meV and $\Gamma_{CM15}=14$meV. Similarly to what is found in \cite{Geisler2012}, the narrower state seems to determine the rate of decay of the photoionization cross-section\\


\section{Concluding remarks}\label{sec:conclusions}

This study represents a significant joint experimental and theoretical endeavor, employing tunable Four-Wave Mixing (FWM) spectroscopy with XUV Attosecond Pulse Trains (APT) and incommensurate IR pulses to time-resolve and assign an autoionizing wave packet within a highly convoluted spectral region of krypton. Experimentally, we successfully identified and probed the coherent electronic superpositions of singly-excited (\sesp{}) and doubly-excited (\fsfp{}) states of krypton launched by the XUV APT. The unambiguous quantum beating of this wavepacket was detected through FWM coupling pathways involving both singly- and doubly-excited states, facilitated by the tunable IR pulses. These results underscore the unique advantages of background-free FWM studies and pave the way for further all-optical investigations into autoionizing electronic wave packets and the fundamental understanding of two-electron processes.

To describe the complex dynamics observed in such a highly correlated system, we developed and proposed a minimal Multi-channel Quantum Defect Theory (MQDT) model. This theoretical framework successfully describes the system's dynamics by treating the excitation to the odd $J=1$ symmetry, including singly-excited (SE) and doubly-excited (DE) states, coupled in a three-channel MQDT model. It further defines two two-channel models, one for each available intermediate reachable symmetries even $J=2$ and $J=0$, to account for intermediate resonances mediating the two-photon IR process. Crucially, this model provides clear mechanisms and sources for the observed features and offers evidence for the interactions between doubly and singly excited autoionizing states of krypton within a specific energy range.

Our model effectively captures the dominant dynamics observed in the experiments. In particular, it is capable of capturing the induced oscillations with a period given by the energy difference between the $7p$ and CM15 states. It also adequately captures the energy dependence of the phase showing an almost exact $\pi$ difference between the signal at the energies of the $5p$ and the CM6 states, which shows the different coupling between singly and doubly excited states. In addition, it also captures the strong dependence of the amplitude of the oscillation with the energy of the IR, highlighting the importance of the intermediate resonances and the degree of resonances of the two photon process. This work also serves as an explicit invitation for future theoretical advancements. More thorough theoretical work is desirable to verify or test the values of the parameters obtained through our current model, as well as to incorporate relativistic effects, which are not yet included. This is particularly important considering that the experimental data, while providing strong constraints, does not necessarily completely determine the value of all parameters. Therefore, the parameters presented here should be regarded as a valuable reference point, encouraging further studies to refine our understanding of these intricate atomic processes.
\begin{acknowledgments}
S.Y-P. and A.S. acknowledge support from the U.S. Department of Energy, Office of Science, Office of Basic Energy Science, under Award No. DE-SC0018251, and CHG under Award No. DE-SC0010545.
\end{acknowledgments}


\bibliography{Refs}
\begin{widetext}
\appendix*
\section{Analytical formula for the position of resonances in two channel case}
For the two channel case, we can obtain simple and instructional analytical formulas for the position of the autoionizing resonances. The formulas acquire their simplest forms when one uses the eigen channel version of multichannel quantum defect theory. This formulation is based on the eigen values $\tan \pi \mu_{\alpha}$ and eigen vectors of $U_{i \alpha}$ of the $K$ matrix. Using the asymptotic forms of these eigen functions one can show that the form of the generalized eigen value problem in this basis is given by:
\begin{equation}
\begin{aligned}
    &\sum_\alpha U_{i\alpha} \sin \pi \mu_\alpha A^\rho_\alpha = \tan \pi \tau_\rho \sum_{\alpha} U_{i\alpha} \cos \pi \mu_\alpha A_\alpha^\rho &\quad \quad \text{if $i$ is open} \\
    &\sum_{\alpha} U_{i\alpha} \sin (\beta_i+\pi\mu_\alpha) A_\alpha = 0 &\quad \quad \text{if $i$ is closed} 
\end{aligned}
\end{equation}
For the case of a two channel system we $U_{11}=U_{22}=\cos \theta$ and $U_{12}=-U_{21}=\sin \theta$. Assuming that the second channel is closes, we can obtain an expression for the first coefficient 
\begin{equation}
    A_1=-\tan \theta \frac{\sin (\beta_1+\pi \mu_2)}{\sin (\beta_1+\pi \mu_2)} A_2
\end{equation}
Replacing in the open channel equation we obtain an expression for the eigen phase shift
\begin{equation}
    \tan \pi \tau = \frac{\sin^2 \theta \sin \pi \mu_1 \sin(\beta_1+\pi \mu_2) + \cos^2 \theta \sin \pi \mu_2 \sin(\beta_1+\pi \mu_1)}{\sin^2 \theta \cos \pi \mu_1 \sin(\beta_1+\pi \mu_2) + \cos^2 \theta \cos \pi \mu_2 \sin(\beta_1+\pi \mu_1)}
\end{equation}
assuming that the the $K$ matrix parameters are not energy dependent one can show that the time delay, $\pi \frac{d}{dE}\tau$, has critical points at
\begin{equation}
    \tan 2 \beta(E_c) = - \frac{\cos^2 \theta \sin 2 \pi \mu_1 + \sin^2 \theta \sin 2 \pi \mu_2}{\cos^2 \theta \cos 2 \pi \mu_1 + \sin^2 \theta \cos 2 \pi \mu_2}
\end{equation}
in order to find the maximum one has to choose the branch of the tangent such that the second derivative of the time delay is negative. This is not an important issue as adding the width constrain at the moment of doing the fit ensures this is not a minimum of the delay. 

In order to relate these parameters with the ones given in the paper, notice that by the definition of the $\mu$ matrix they are related by
\begin{equation}
    \mu_{ij} = \sum_\alpha U_{i \alpha} \mu_\alpha U_{j\alpha}
\end{equation}

Another formulation, that is perhaps more transparent and directly expressible in terms of the $K$ matrix, is the one inspired by the channel elimination procedure from Chapter 8 in \cite{fanoRau2012atomic}. 
Let channel 1 be open and channel 2 be closed. Using Eq.~8.62, we can ``eliminate'' the open channel (imposing purely outgoing wave boundary conditions in channel 1) to get the purely closed channel effective K-matrix which is now non-Hermitian, reflecting the fact that the energies between the two thresholds are complex, i.e. $E_r - i \Gamma/2$.
\begin{equation}
    K_{22}^{\rm eff}= K_{22}+i \frac{K_{21} K_{12}}{1-i K_{11}} \equiv \tan{\pi \tau}
\end{equation}
And now the complex energies of the autoionizing states have the following Rydberg formula:
\begin{equation}
    E_n=IP_2 -\frac{1}{2 \nu_n^2}
\end{equation}
where $\nu_n= n-\tau$.  If the original 2$\times 2$ $K$-matrix is energy-dependent, then $\tau=\tau(E)$ is energy-dependent, of course, and solving for the complex energies $E_n$ would involve demanding self-consistency between Eq.(2) and $\nu_n= n-\tau(E_n)$ .  To break this quantum defect down into real and imaginary parts, we can simplify a little bit farther, namely:
\begin{equation}
    \tau_{\rm re} +i \tau_{\rm im} = \frac{1}{\pi} \arctan{{\biggr [} (K_{22}-\frac{K_{11}K_{12}^2}{1+K_{11}^2})+i \frac{K_{12}^2}{1+K_{11}^2}    {\biggr ]}}.
\end{equation}
The same type of analysis works for a 3-channel system, e.g. if there are two channels open and one closed, one can eliminate the open channels, i.e. implement Fano-Rau Eq.8.62, and again express the complex quantum defect of resonances in the closed channel in terms of the K-matrix elements.  For the 2-channel case discussed in the text, this gives an energy-independent closed-channel quantum defect equal to $\tau_{5s}=2.158+0.110 i$, and the $n=5s$ resonance position is at $E_{5s}=25.836$ eV with width $\Gamma_{5s}=0.259$ eV.  Instead, the complex quantum defect for the $nd$ series is $\tau_{4d}=1.238+0.0485 i$ giving a 4d level position $E_{4d}=25.731$ eV and width $\Gamma_{4d}=0.125$ eV.

\section{One photon photionization amplitude}\label{app:B}
In order to describe the situation explored in \cite{Geisler2012} with our current model we need to derive a formula for the photoionization probability. We again employ perturbation theory to determine the amplitude. Using he first order term of the time propagator expansion we find a term given by
\begin{equation}
    C^J_i(E) = \mathcal{F}_o \gamma \sqrt{\pi}\int d\delta c(\delta) \exp \left[-\left(\gamma \frac{E-\delta-\omega}{4}\right)^2 \right] e^{i(E-\delta)t_o} \bra{\psi^{J,-}_i(E)}z\ket{\psi^{1,-}(\delta)}
\end{equation}
where we find the amplitude for both open channels and both possible angular momentum values, and then add them incoherently to obtain a probability
\begin{equation}
    P(E)=\sum_{J=0,2}\sum_{i=1,2} |C_i^{J}(E)|^2
\end{equation}

\end{widetext}

\end{document}